\documentstyle[twocolumn,prb,aps,psfig]{revtex}

\begin{document}
\draft{}

\title{Electron Spin Relaxation under Drift in GaAs}

\author{E. A. Barry, A. A. Kiselev, and K. W. Kim}
\address{Department of Electrical and Computer Engineering,
North Carolina State University, Raleigh, North
Carolina 27695-7911}


\maketitle


\begin{abstract}

Based on a Monte Carlo method, we investigate the influence of transport 
conditions on the electron spin relaxation in GaAs.  The decay of initial 
electron spin polarization is calculated as a function of distance under the 
presence of moderate drift fields and/or non-zero injection energies. 
For relatively low fields (a couple of $kV/cm$), a substantial amount of 
spin polarization is preserved for several microns at 300 K. However, it is
also found that the spin relaxation rate increases rapidly with the drift
field, scaling as the square of the electron wavevector in the direction of the
field.  When the electrons are injected with a high energy, a pronounced 
decrease is observed in the spin relaxation length due to an initial increase 
in the spin precession frequency.  Hence, high-field or high-energy transport
conditions may not be desirable for spin-based devices.

\end{abstract}

Recently, there has been an increasing interest in the emerging field
of spin electronics (spintronics), where the electron's spin degree of 
freedom is exploited.~\cite{Prinz,Wolf,rashba_chall}  Many devices and 
applications, including giant-magnetoresistive structures~\cite{daughton_gmr}
and magnetic random access memories,~\cite{comstock_mram} have been suggested
with varying degrees of success.  Furthermore, the suggestion of utilizing
electron spin for quantum information processing is a natural extension of 
spintronics since the electron provides an ideal condition for a qubit.

Of the potential applications, the hybrid devices~\cite{Datta} that 
combine "traditional" semiconductor electronics with the utilization of the 
spin state are currently at the center of attention for their increased 
functionality and the ease of integration.  Electron spin states in
semiconductor structures relax (depolarize) by scattering with imperfections 
or elementary excitations such as other carriers and phonons.  Therefore, to 
realize any useful spintronic devices, it is essential to understand
and have control over spin relaxation such that the information is not 
lost before a required operation is completed.  So far, most of the studies on 
spin relaxation have been focused on electrons with a thermal or near-thermal
distribution.~\cite{piku}  However, electron spins in the spin-based devices
can be subject to highly non-thermal transport conditions including high drift 
fields for high-speed transfer of spin information.  

In this work, we investigate the influence of transport conditions on electron
spin relaxation in n-type GaAs.  A Monte Carlo approach is used to simulate 
electron transport, including the evolution of spin polarization,~\cite{kiselev}
to determine spin relaxation lengths and times.  A non-parabolic energy-band 
model (with $\Gamma-L-X$ valleys) is used
along with the relevant momentum relaxation mechanisms such as polar optical 
and acoustic phonon deformation potential scattering.  Of the three main spin 
relaxation mechanisms [i.e., Bir-Aronov-Pikus, Elliot-Yafet, and 
D'yakonov-Perel' (DP) mechanisms],~\cite{piku,fabian_relax,PilHun}  we only 
consider the DP processes since this is the dominant mechanism in the regime
of interest, namely GaAs at 300K.  It is also important to note that the
current study is limited to the low energy cases, such that electrons 
are found only in the ${\Gamma}$ valley. This is due to the lack of materials
parameters on spin-orbit spitting in higher energy valleys.

The DP Hamiltonian due to spin-orbital splitting of the conduction band
may be written as~\cite{Dyakonov}

\begin{equation} \label{crit-1}
	H=\frac{\hbar}{2} \, \vec{\sigma} {\bf \cdot} \vec{\Omega}_{eff};
\end{equation}
and
\begin{equation} \label{crit-2}
	\vec{\Omega}_{eff}=\frac{\alpha \hbar^{2}}{m\sqrt{2 m E_{g}}}[k_{x}(k_{y}^{2}-k_{z}^{2})\hat{x}+\text{cyclic perm.}] \,;
\end{equation}
where $\alpha$ is a dimensionless, material-specific parameter which gives the 
magnitude of the spin-orbit splitting $\alpha \simeq
4\eta \:m/\sqrt{3-\eta} m_0$ and $\eta=\Delta/(E_g+\Delta)$, $m$ 
is the effective mass, $\vec{k}$ is the electron wave vector, $E_{g}$ is the 
energy separation between the conduction band and valence band at the $\Gamma$
point, and $\Delta$ is the spin-orbit splitting of the valence band. The material 
parameters for GaAs are listed in Table I.

The quantum mechanical description of the evolution of the spin 1/2
is equivalent to the evolution of the classical momentum $\vec{S}$ under an 
effective magnetic field $\vec{\Omega}_{eff}$ with the equation of motion

\begin{equation} \label{crit-4}
	\frac{d\vec{S}}{dt}= \vec{\Omega}_{eff} \times \vec{S} ~.
\end{equation}

\begin{table}
\begin{center}
\begin{tabular}{ ccccccc}    
	$m/m_0$ 	& $\hbar \omega_0 $ 	&  $E_l$   & $E_g$   & $\alpha_{np}$ &   $\Delta$		\\
			& 	$eV$		&  $eV$	   & $eV$    &		     &		$eV$		\\
	\hline
 	$0.067$  	&  $0.0354$ $ $		&  $7.0$   & $ $ $1.424$ $ $& $0.616$ $ $    & 	 $0.341$ $ $    \\			
 
\end{tabular}
\caption{Material Parameters for GaAs. $m/m_0$ is the $\Gamma$ valley effective mass ratio, $\hbar \omega_0$ is the
optical phonon energy, $E_l$ is the acoustic phonon deformation potential,  $E_g$ is the energy band gap separation at the 
$\Gamma$ point, $\alpha_{np}$ is the non-parabolicity factor in the $\Gamma$ valley,  and $\Delta$ is the spin-orbit 
splitting of the valence band.}
\end{center}
\end{table}

\begin{figure}
\begin{center}
\psfig{figure=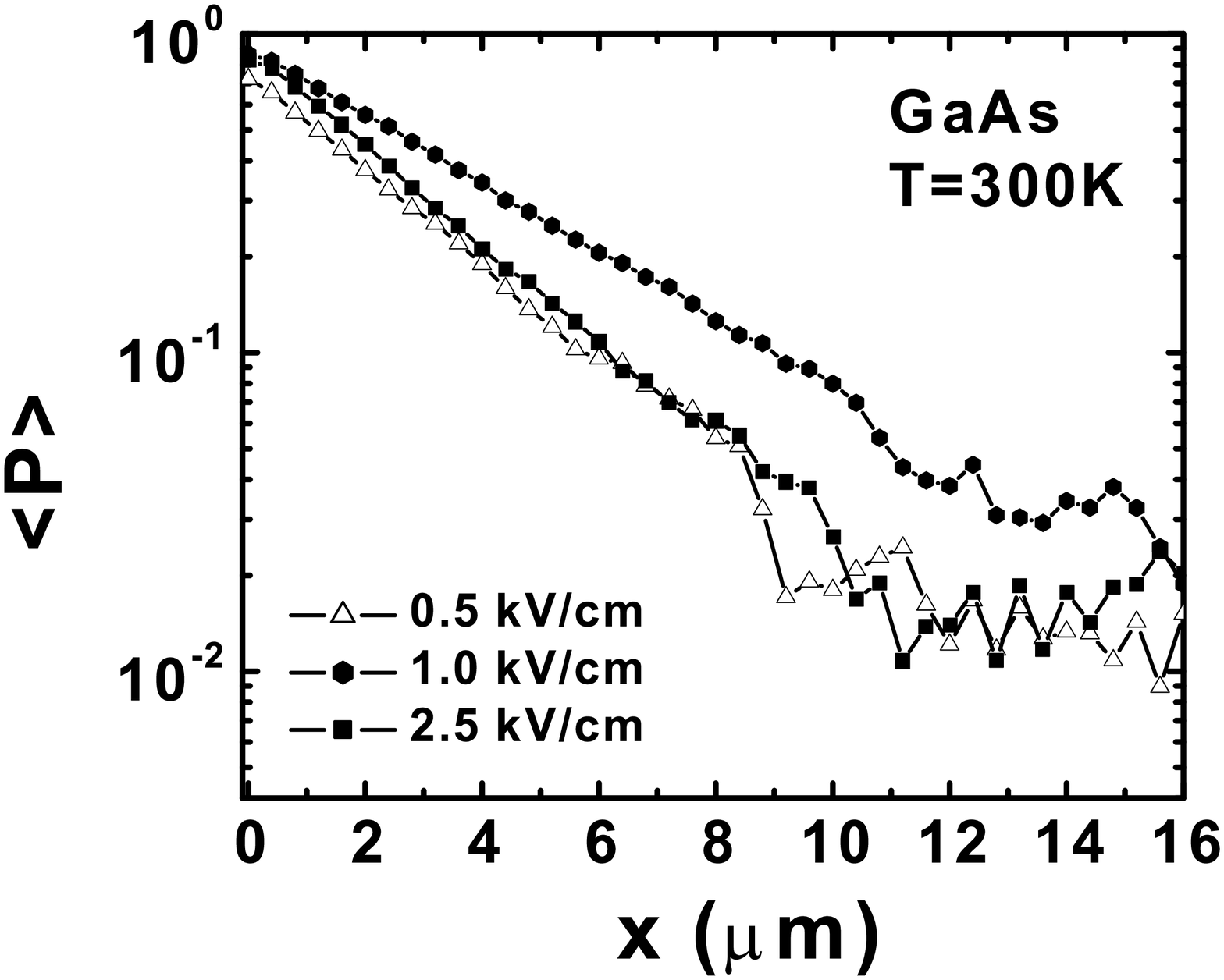,width=8.5cm,height=8.0cm,angle=0}
\caption{\label{f:f-1}Polarization $<P>$ of injected electrons as a function of distance.}
\end{center}
\end{figure}

Figure~\ref{f:f-1} depicts the average polarization $<P>$ of electrons calculated
as a function of distance in GaAs.  The simulation follows the evolution of 
$1 \times 10^{4}$ spin-polarized electrons from an injection plane.  Although 
all electrons are initially polarized ($<P_0>=1$), polarization at $x=0$ in 
this figure is not exactly unity due to the averaging over a finite mesh size.
Clearly, $<P>$ decays nearly exponentially with an order of magnitude drop 
(to 0.1) in approximately 4 $\mu m $ to 7 $\mu m $ (5.5 $\mu m $ to 9 $\mu m $ 
when the non-parabolicity is not considered).  

The observed dependence of decay
on the applied electric field is non-monotonous with the result of 2.5 $ kV/cm$
placed between the two cases with smaller fields.  This can be explained by 
the two competing factors that contribute to the spin relaxation distance. 
As the field becomes larger in the linear regime, the electron momentum and the
drift velocity increase in the direction of the field.  On the other hand, the 
increased electron momentum also brings about a stronger effective magnetic 
field as shown in Eq.~(\ref{crit-2}).  Consequently, the electron precession 
frequency becomes higher, resulting in faster spin relaxation 
(i.e., shorter spin relaxation 
time).  The interplay of these opposing trends (i.e., higher drift velocity and
shorter relaxation time) determines the observed behavior of spin relaxation 
length as a function of electric field.

To examine the dependence of spin relaxation in detail, we calculate
the spin relaxation rate $\tau_s^{-1}$ as a function of electric field.
The value of $\tau_s$ is estimated by approximating the spin decay to be
exponentially dependent on length. The slope of Fig.~\ref{f:f-1} may 
be taken to be a characteristic length $L_D$, such that

\begin{equation} \label{crit-5}
	<P>= A \exp{(-L/L_D )};
\end{equation}

\begin{figure}
\begin{center}
\psfig{figure=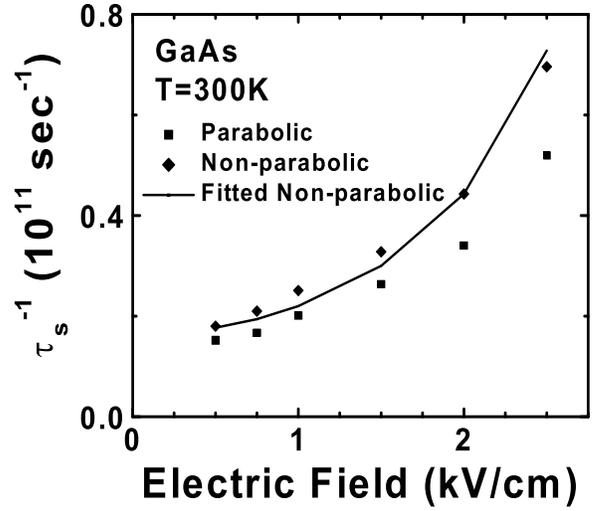,width=8.5cm,height=8.0cm,angle=0}
\caption{\label{f:f-2}Spin relaxation rate $\tau_s^{-1}$ as a function of electric 
field $F$.  The data points are the results calculated with or without the
non-parabolicity.  The solid line represents the $ < k_x^2 >$ scaling for the
non-parabolic band.}
\end{center}
\end{figure}

\noindent
where $A$ is a normalization factor. The spin relaxation rate may then be 
computed as $\tau_s^{-1}=v_{dr}/L_D$, where $v_{dr}$ is the average drift 
velocity.  For comparison, the results for the parabolic energy band are 
shown in Fig.~\ref{f:f-2} along with those with the non-parabolic band.

As expected, the spin relaxation rate increases rapidly with the applied 
electric field, which is also in accord with a recent experimental 
observation.~\cite{ohno} Although $\tau_s$ for low electric fields is nearly 
the same in both the parabolic and non-parabolic band models, the spin relaxation 
rate becomes considerably larger in the non-parabolic case for fields higher
than approx. 1.5 $kV/cm$. The values for $\tau_s$ 
at the low fields agree well with those obtained for near thermal 
electrons.~\cite{PilHun}

The dependence of spin relaxation on electric field can be best understood
by considering the functional form of the effective magnetic field in 
Eq.~(\ref{crit-2}).
When the applied electric field is along the $ x $ direction, the interaction
Hamiltonian reduces approximately to
\begin{equation} \label{crit-6}
	H \propto (k_z \hat{z}-k_y \hat{y}) k_x^2 ~.
\end{equation}

\noindent
Since the average values of $ k_y $ and $ k_z$ remain small, it is reasonable 
to assume that $\tau_s^{-1}$ will scale as $<k_x^2>$.  The solid line in 
Fig.\ 2 provides the $<k_x^2>$ scaling for the non-parabolic band, which 
clearly illustrates good agreement over the entire field range under 
consideration. This finding indicates that the common assumption of electron 
temperature $T $ to the third power dependence 
[i.e., $ 1/{\tau^{DP}_{s}} \propto (k_{B}T)^3 \tau_p $, where $\tau_p$ is the 
momentum relaxation time]~\cite{piku,PilHun}
substantially overestimates the spin relaxation rates in the drift regime.
Considering the nature 

\begin{figure}
\begin{center}
\psfig{figure=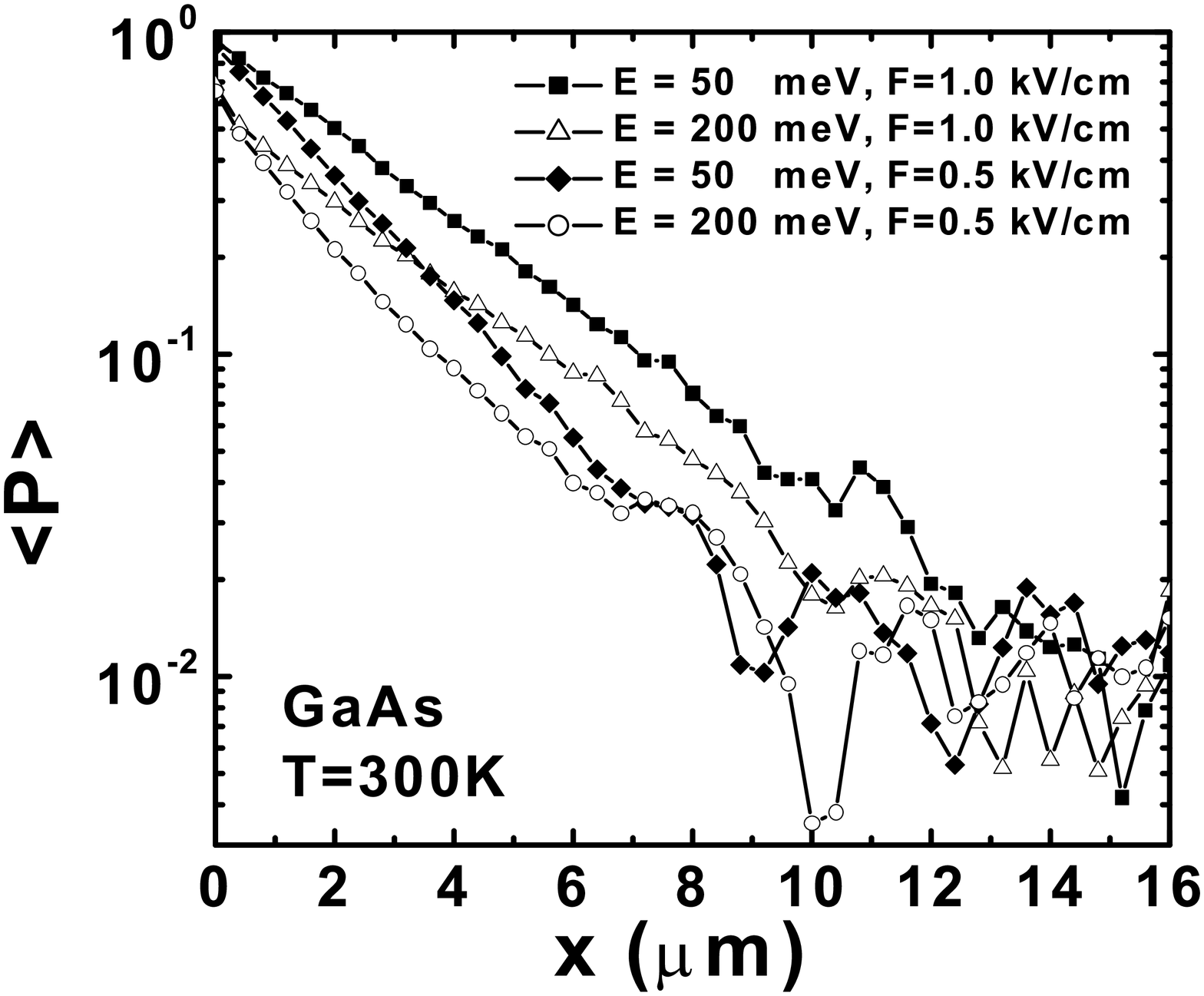,width=8.5cm,height=8.0cm,angle=0}
\caption{\label{f:f-3}Electron polarization $<P>$ vs. distance as a function of injection
energy and applied electric field.}
\end{center}
\end{figure} 

\noindent
of drift in this case, the $T^2$ scaling may be more 
appropriate. 

However, the applicability of this scaling rule is also limited. 
As the electron temperature increases, it is clear that one needs to
take into account the effect of non-parabolicity 
on the spin-orbit splitting (i.e., DP Hamiltonian) itself.
By ignoring the higher band contributions, Pikus and Titkov 
obtained the following estimation in place of $ \alpha m^{-3/2}$ 
in Eq.~(\ref{crit-2}) with the rest of the terms unchanged:~\cite{piku} 

\begin{equation} \label{crit-3}
  \alpha(E) m^{-3/2}(E) =\alpha m^{-3/2}\left(1-\frac{E(\vec{k})}{E_g}\frac{9-7\eta+2\eta^2}{3-\eta}\right).
\end{equation}
\noindent
At the constant electric field of 2.5 $kV/cm$, Eq.~(\ref{crit-3}) results in
$\alpha(E) m^{-3/2}(E)$ that is smaller than $\alpha m^{-3/2}$ by 
approximately 10\%.  The deviation shrinks as the electric field 
decreases,\cite{lowE} leading to a minor change in the $T^2$ scaling 
of $\tau_s^{-1}$. 
For the fields beyond the range currently under consideration (0.5$-$2.5 
$kV/cm$), the influence of the additional term in Eq.~(\ref{crit-3}) can 
become crucial.

In addition to the influence of the drift field, we also investigate spin 
transport and relaxation with varying injection energies. 
As mentioned earlier, one of the fundamental challenges in 
spintronics has been the electrical injection of spin polarized current into 
a semiconducting material. Noting recent suggestions~\cite{Rashba_inj} 
for the utilization of tunnel injection, we may anticipate relatively high 
electron energies at the injection point.

In Fig.~\ref{f:f-3} we plot the results of simulations where the injection energy 
and applied electric field are changed.  It can be seen that the electrons 
quickly lose the extra energy and their momentum distribution stablizes within 
1 $\mu m$ or so, even in the cases of higher energy injection (200 $meV$). 
Therefore, we have two distinct slopes. The first is a more rapid decay
in the non-local transport region, corresponding to a large $\vec{k}$ and 
subsequently a large precession vector $\vec{\Omega}_{eff}$. 
Once thermalization occurs, the slope becomes smaller and eventually coincides 
with those shown in Fig.~\ref{f:f-1} with corresponding electric fields.
For the 200 $ meV$ case with an applied field of 1 $kV/cm$, the reduction in 
depolarization length is fairly significant, approximately 20 \%.

In summary, we have investigated spin-dependent transport in GaAs in the
drift regime.   Specifically, we calculated the spin relaxation time and 
characteristic decay lengths based on a Monte Carlo model.  The decay of 
electron spin polarization is found to be nearly exponential and
the length at which spin relaxation/depolarization occurs was found to be 
relatively large, on the order of 5$-$10 $\mu m $. 
It is also found that the spin relaxation depends strongly on the drift 
conditions, scaling as the square of the electron wavevector in the direction 
of the field. Hence, the commonly assumed $T^3$ scaling is not applicable
in the drift regime. When the electrons are injected with a high-energy, 
a significant reduction is observed in the spin relaxation length.
Our calculation results are in good agreement with the data available
in the literature.

This work was supported by the Office of Naval Research and the Defense 
Advanced Research Projects Agency.  The authors are grateful to Y. Semenov 
for helpful discussions.

\end{document}